# A Stream Pipeline Framework for Digital Payment Programming based on Smart Contracts

Zijia Meng[1,a*]
Grade 12
Wuhan Britain-China School
Wuhan, Hubei, China
Zeejamengzijia@outlook.com

Victor Feng[2,b]
Engineering Dept,
Unizon Blockchain Technology Pty Ltd
Sydney, NSW, Australia
victor@unizon.au

*Abstract*—Digital payments play a pivotal role in the burgeoning digital economy. Moving forward, the enhancement of digital payment systems necessitates programmability, going beyond just efficiency and convenience, to meet the evolving needs and complexities. Smart contract platforms like Central Bank Digital Currency (CBDC) networks and blockchains support programmable digital payments. However, the prevailing paradigm of programming payment logics involves coding smart contracts with programming languages, leading to high costs and significant security challenges. A novel and versatile method for payment programming on DLTs was presented in this paper – transforming digital currencies into token streams, then pipelining smart contracts to authorize, aggregate, lock, direct, and dispatch these streams efficiently from source to target accounts. By utilizing a small set of configurable templates, a few specialized smart contracts could be generated, and support most of payment logics through configuring and composing them. This approach could substantially reduce the cost of payment programming and enhance security, self-enforcement, adaptability, and controllability, thus hold the potential to become an essential component in the infrastructure of digital economy.

*Keywords - Blockchain, Distributed Ledger Technology, Smart Contract, Programmable Payment, Stream Pipeline*

I. INTRODUCTION

Secure and self-enforced payments with programmable digital currencies play the central role in the future of digital economy. Researchers in monetary authorities assert that the application of programmable digital money can vastly enhance the integration and interoperability of international trading and economic cooperation. [1] For governments, central banks, and financial regulators, programmable money provides a novice economic and monetary governance tool that enables unprecedented transparency, efficiency, and controllability. One of the most prominent value propositions of the programmable money is payment programming [2][3]. Programmable payments are payments that are self-enforced, automatically executed when certain conditions are met. Thus, these payments are automated and follow an inherent, predetermined rules. Ethereum, for example, introduces smart contracts that directly manipulate tokens, digital value records stored in the blockchain. In public blockchains like Ethereum, every user can create and deploy a smart contract, effectively controlling and automating payment processes.

Programming payments is an immensely valuable feature that holds the potential to become one of the most pervasive applications of blockchain and the Distributed Ledger Technology (DLT). Let's explore a few examples illustrating the power of programmable payments:

- Authorized Payment: Users pre-approve a transaction and delegate another user to execute payments within a specific expense limit.

- Cascading Payment: Several recipients create a payment chain, ensuring that when a payment is initiated, funds circulate through each recipient sequentially.

- Recurrent Payment: Users set up scheduled payment plans, ensuring the automated transfer of either a fixed or variable amount of funds at predefined intervals throughout the specified payment term.

- Pay-to-many: Users execute payments to multiple recipients simultaneously. The funds are distributed to each recipient based on a predefined set of distribution rules, streamlining the process and enhancing efficiency.

It's evident that these payment patterns find widespread applications in real-world scenarios, both in daily life and business operations. However, few users have embraced programming their payments. The question arises: Why hasn't this highly advantageous innovation been widely adopted?

The primary challenge lies in the intricate nature of secure smart contract programming. Given its involvement with value, smart contract programming is notoriously complex. With numerous publicized incidents of value losses and breaches, it's universally acknowledged that even a minor security vulnerability in a smart contract could result in significant financial losses, often amounting to millions of dollars. A seasoned smart contract programmer needs to invest thousands of hours sharpening their secure coding skills and bug hunting abilities. Furthermore, their team must allocate a substantial budget for code auditing. The overall expenses can be overwhelming for a typical institutional user, let alone an individual.

It is evident that there needs to be a cost-effective, user-friendly, and convenient method for secure payment programming for it to gain traction and be widely adopted among end users.

II. AN OVERVIEW OF EXISTING METHODS

There are some efforts being made to simplify smart contract programming.

*2.1 Code libraries*

Providing code libraries is the most widely adopted solution. Notably, open-source libraries like OpenZeppelin [4] offer a variety of audited, pre-built, and battle-tested code modules that are readily accessible to smart contract developers. Importing these modules into the smart contracts under development is convenient. The teams behind these code libraries are among the most experienced in secure contract programming. These libraries, extensively tested and utilized by hundreds of thousands of developers, are considered some of the most trusted pieces of code across the industry.

However, these code libraries primarily offer fundamental functions in order to cater to a broad spectrum of smart contract programming demands. Consequently, they are predominantly perceived as general-purpose tools tailored for professional developers, rather than being conducive to end users engaged in programmable payments. Despite the extensive experiences a developer may have, secure contract programming remains exceptionally intricate to him.

*2.2 Programmable money enhancement*

the Monetary Authority of Singapore (MAS) introduces an innovative protocol for implementing programmable money called Purpose Bound Money (PBM) [5]. This protocol aims to improve settlement efficiency, user experience, and merchant onboarding across various ledger technologies and use cases. Central to the PBM protocol is the programming logic that enables the customization of digital money. This results in an easier approach of programming payments.

PBM offers a robust framework for programming payments by governing its utilization through bespoke conditions. This allows for an unprecedented level of customization in how money flows in digital networks. Yet, it is critical to note that PBM, while revolutionary, is not a direct tool for end users; rather, it serves as a platform upon which developers and institutions can build more user-friendly and adaptive payment systems.

*2.3 Intent-Centric architecture*

One of the recent promising developments in the DLT space is the Intent-Centric architecture [6], spearheaded by Paradigm, a leading Web3 investment firm. In the long run, the architecture has ambitious plans to incorporate Artificial Intelligence, allowing these systems to interpret and execute intricate user intents, thereby automating complex transactions across a range of smart contracts or DeFi protocols. While this architecture holds the potential to substantially simplify payment programming for end users, widespread adoption still appears to be a few years away.

It is evident that the above existing solutions aimed to simplify payment programming fall short in addressing the needs of today's users. Traditional methods, while technically advanced, often present barriers of complexity that make them inaccessible to the average user. What is urgently needed is a novel, pragmatic approach that can be implemented immediately and is user-friendly enough to be accessible to a broader audience. This new paradigm should not only streamline the processes of programming payments but also lower the technical barriers to entry, enabling common users to do the work with ease and security.

III. STREAM-ORIENTED PROGRAMMING AS A PARADIGM

It does not require acute insight to notice that the pattern of payment programming closely echoes that of text processing scripting in Unix-like operation systems. In Unix "shell-scripting", a popular method for achieving a text processing goal is to serialize the text to construct a character stream that "flows" through a pipeline of sequential commands [7]. Each command in the pipeline transforms the stream in a specific manner. Cumulatively, these transformations generate desired output at the end of the process. This paradigm of text programming is known as command pipeline or stream-oriented text processing.

A complex example showed in Figure 1 finds the 10 most frequently occurring words in a file.

```
cat  filename.txt  |  tr  -s  ' '  '\n'  |
tr'[:upper:]' '[:lower:]' | sort | uniq -c |
sort -nr | head -n 10
```

Figure 1. An example of text processing pipeline

In this example, command "cat" reads the file and constructs the text stream, which is directed by the pipeline operator '|' to the standard input of the first "tr" command. This "tr" command replaces spaces with newlines to put each word on a new line, then pipelines to the next "tr '[:upper:]' '[:lower:]'" command, which converts all characters to lowercase. The text stream then goes to "sort" and "uniq" commands in turn, which first sorts the words then counts the unique occurrences. The last two steps are "sort" and "head". "sort -nr" command sorts these counts in descending numerical order, then at last the "head -n 10" command display the top 10 most frequent words on the screen. The same text stream flows through the seven commands and gets transformed one small step a time. As the result, a complex text processing task is accomplished by flowing the text stream through these simple commands pipelined together.

It has become evident that the stream-oriented programming paradigm extends well beyond text processing. For instance, in the renowned MapReduce framework, Google utilizes this paradigm to achieve massive parallelism in big data processing on large clusters, establishing it as the de facto approach in the domain [8].

If a programming task can be abstracted such that a single object undergoes sequential, non-recursive procedures, the stream-oriented paradigm often emerges as a more straightforward and efficient method for accomplishing the task.

It seems that payment programming aptly falls into this category. At its core, payment programming is fundamentally about creating a money stream that originates from source accounts and flows to target accounts, all while adhering to a specific set of rules or conditions. Therefore, payment programming in smart contract enabled DLT environments can be modeled using a stream-oriented method with the following steps:

1. Represent money as token streams;

2. Define a standardized structure for stream-processing smart contracts that receive incoming token streams, transmit outgoing token streams, and handle the messages and errors in a normalized manner.

3. Engineer a collection of foundational yet configurable smart contract templates to serve as the essential building blocks for payment programming;

4. Develop a tool that empowers users to customize and deploy smart contracts through the configuration and instantiation of these templates.
5. Devise a systematic method for chaining and pipelining multiple such smart contracts to fulfill the payment programming objectives.

The stream-oriented paradigm for payment programming offers a host of advantages that address the complexities and limitations commonly associated with traditional methods. By representing money as tokenized streams and employing a standardized architecture for smart contracts, we facilitate a more intuitive, efficient, and robust framework for payment processing.

## VI. A MODULAR FRAMEWORK FOR STREAM-ORIENTED PAYMENT PROGRAMMING

The proposed framework consists of three core types of node component: the Stream Originator, the Router, and the Endpoint. The Stream Originator is a specialized class of smart contracts tasked with converting raw fungible tokens into token streams. On the other hand, the Router represents another distinct class of smart contracts, specifically designed to execute certain functions, and forward these token streams to subsequent nodes in accordance with predefined rules. The Endpoint contracts perform the reverse operations to transform streams back to plan token payments.

To expedite the development of these nodes, the framework will offer a curated library of smart contract templates. Users can then effortlessly choose from these templates to configure and instantiate the nodes they require.

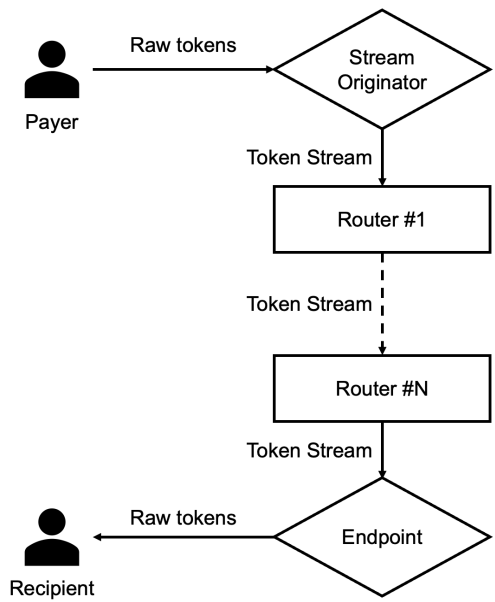

Figure 2. A stream pipeline comprised of N Routers

### 4.1 Stream Originator

The first core component is Stream Originator. Its primary role is to covert raw token flows into token streams.

In the realm of DLT, digital money is predominantly represented through the raw fungible token standard ERC-20 [9] and its derivatives. However, the ERC-20 standard falls short in qualifying as a "stream" for two key reasons:

First, the ERC-20 standard lacks a notification mechanism to alert recipient smart contracts when a token transfer occurs.

Second, the ERC-20 standard also falls short in providing a standardized protocol for error handling, which is crucial in stream-oriented programming.

Thus, we require a mechanism to convert basic token flows into a stream equipped with inherent notification and error-handling protocols.

Stream Originator serves as a specialized smart contract that functions as the initiation point for stream-oriented payment. Its primary duty is to convert raw tokens into a token stream and establish the protocols that govern the entire stream. These protocols encompass:

- Connection Protocol: Every node sets up one or more incoming and outgoing connections with other nodes.
- Fund Transfer Protocol: Each node dispatches funds to succeeding nodes by invoking their `onStreamReceived()` method. In response, the recipient nodes utilize the `transferFrom()` method to execute the fund transfer.
- Event notifying protocol: After a successful payment, the originating node must trigger a `Sent` event.
- Error handling protocol: if an error arises, the node needs to communicate the issue by firing the `StreamError` event. Based on the severity of the error, the node can decide whether to proceed with the payment, hold onto the funds, issue a refund, or direct the funds to designated accounts.

### 4.2 Router

Routers plays the central role in stream-oriented payments. Like the routers in the Internet, a Router in stream-oriented programming has an incoming connection and one or more outgoing connections. It receives the token stream from the incoming connection, then either retains tokens or forward the them onward through the outgoing connections, either to the subsequent Router or an Endpoint. However, a Router isn't merely a passive conduit; it's equipped to execute a series of logics and actions in the interim between receiving and forwarding. For instance, a Router may:

- Report the payment status to a logging system;
- Lock the tokens for a specified duration.
- Check if the payment conforms to regulations before passing on the funds.
- Assess a predefined set of rules to determine both the destination and amount of the fund transfer;
- Accrue the funds and execute recurring payments based on the predefined rules;
- Establish checkpoints for human intervention or arbitration, such as in the context like escrow payments.
- Remit taxes or fees simultaneously.
- Allocate payments among multiple parties according to established or agreed criteria, etc.

Every Router must adhere to the protocols established by the Stream Originator of its respective pipeline. Specifically:

- It must implement the `onStreamReceived()` method to execute the logics and receive or reject the token stream;
- It is required to invoke the `onStreamReceived()` method of the subsequent Routers or Endpoints;
- Upon completing the fund transfer, it must trigger the `Sent` event;
- Errors must be handled appropriately.

Thus, a Router must be "programmable." In essence, designing a series of specialized small Routers and linking them can be more straightforward than crafting a comprehensive, monolithic smart contract with equivalent functions.

*4.3 Endpoint*

An Endpoint serves as the concluding node in a stream payment pipeline. Its primary function is to convert the received token stream back into plain tokens, depositing them directly into the accounts of the final recipients. Alternatively, it can hold onto these tokens, allowing recipients to claim them later.

Every Endpoint must adhere to the protocols established by the Stream Originator of its corresponding pipeline, including:

- It must implement the `onStreamReceived()` method, receiving the token stream and resetting it to be plain tokens;
- Errors must be handled appropriately.

The Stream Originator and Endpoint are highly standard and can work in most scenarios without any changes. However, users need to configure Routers to implement specific payment logics.

*4.4 Router Template Library*

To construct a specific payment logic often requires multiple Routers each with unique configurations. To create these Routers from scratch adhering high standard of security each time would be immensely time and labor-intensive. A Router template library would significantly streamline this process, offering pre-programmed Router templates designed to cater to a vast array of payment logics.

Creating a comprehensive collection of Router Templates to cover every conceivable complex payment scenario would incur prohibitive costs. However, it is reasonable to assert that a limited set of Router Templates could accommodate most common payment logics.

Here are some proposed Router Templates that the library should include:

- Reporting Router: This template seamlessly forwards the token stream while simultaneously triggering events to report the status of the stream.
- Time-locking Router: This variant holds incoming tokens and locks them within the contract, later releasing them to the next node according to a predetermined time schedule. It should support complex releasing schemes such as recurring releases.
- Threshold Router: This template accumulates incoming tokens until the total reaches a predetermined threshold, at which point it forwards the token stream to the next node.
- Distributing Router: This template can be configured to distribute received tokens to multiple nodes according to predefined distribution rules.
- Conditional Router: This variant evaluates a set of predefined conditions before forwarding the token stream to the next node.
- Oracle-Directed Router: This template retains the received tokens while awaiting instructions from trusted oracles to determine the destinations and amounts for token stream forwarding.
- Waterfall Router: This template organizes destination nodes into multiple tiers, prioritizing the distribution of received tokens based on the hierarchy established through configured rules. Accordingly, higher-tier nodes are fully compensated before any funds are directed towards lower-tier nodes, ensuring a structured and staged fund distribution process.
- Goalkeeper Router: This utility template serves as a default error handler. In instances where a Router is unable to process an error, it redirects the funds and accompanying messages to this Router. The Goalkeeper Router then manages the error and funds, following a predetermined strategy aligned with the details conveyed through the received messages.

Most payment scenarios should be addressed by skillfully configuring and combining these Router templates. For certain special cases that are difficult to accommodate with existing templates, users may need to develop tailor-made Routers to facilitate the processes.

V. BENCHMARK AND EVALUATION

To assess the effectiveness of the proposed framework, we initiated an experiment comparing the stream pipeline solution with the traditional monolithic solution. Given that the primary value propositions of the framework revolve around development efficiency and security auditing, we identified the following three factors as key indicators:

1. Lines of Code (LOC) – gauge the extent of development workload necessitated for crafting a payment logic.
2. Estimated Auditing Cost – ascertained based on the prevailing market rate.
3. Gas consumption - to assess how the stream pipeline solution impacts the transactional costs. [10]

The payment scenario tested involves periodically paying a group of individuals and reporting the payment details to the tax authority. To construct such a payment logic, the stream pipeline is composed of:

1. The Stream Originator, which acts as the initial repository for deposits and retains the funds.
2. A Time Locking Router set to periodically extract funds from the Stream Originator.

3. A Distributing Router that receives the funds from the previous Router and distribute them to various subsequent Routers based on a predefined allocation plan.
4. Multiple Reporting Routers, each tasked with receiving funds, notifying the tax authority of the income specifics, and subsequently directing the funds to its designated Endpoint.
5. Multiple Endpoints, which are responsible for ensuring the accurate transfer of funds into the final recipient's accounts.

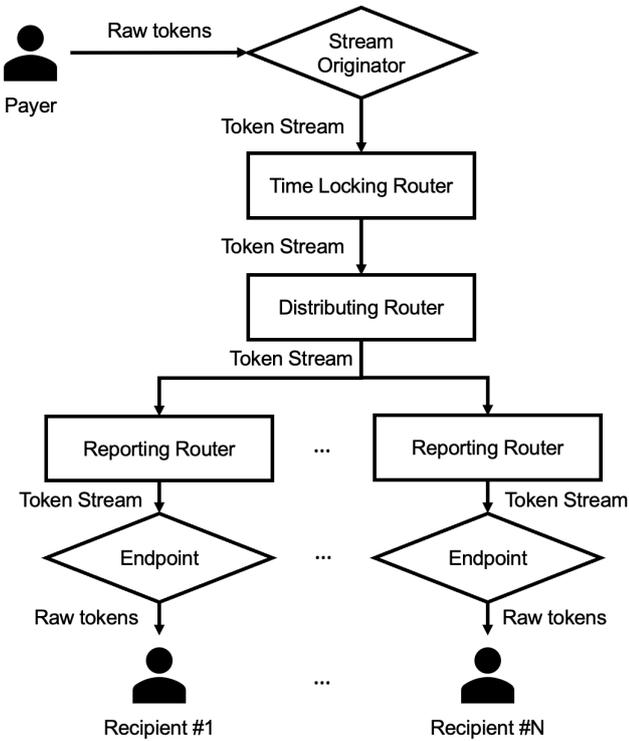

Figure 3. The stream pipeline for the evaluation

To benchmark this solution, one of the authors developed two contracts in parallel: the first being a monolithic contract encompassing all functionalities within a single framework, and the second leveraging the Stream Pipeline solution grounded in the Router Template Library. Subsequently, we juxtaposed these two solutions, deriving the following results:

TABLE 1. THE BENCHMARK OF STREAM PIPELINE PAYMENT SOLUTION

|  | **Monolithic Contract** | **Stream Pipeline** |
|---|---|---|
| LOC | 1,175 | 113 |
| Auditing Quotation | $8,000-$9,000 | $1,000 |
| Gas Consumption | 257,874 | 549,995 |

The results in Table 1 illustrate that, compared to the traditional method, the stream pipeline solution considerably diminishes the development workload, consequently driving down audit costs significantly. However, it demands 113.3% more gas than the monolithic solution, indicating a potential drawback, especially in scenarios where gas prices surge notably, potentially mitigating the advantages offered by this novel solution.

In addition to increased gas consumption, there are two other significant limitations that should not be overlooked:

1. It is fundamentally challenging to incorporate loops into the pipeline.
2. The stream pipeline segments a payment process into multiple stages, thus disabling "flash loan" transactions [11], a noted "innovative" feature in Decentralized Finance (DeFi).

VI. CONCLUSION

Despite certain limitations, the stream-oriented payment programming framework presents a competitive solution in the evolving financial landscape. The inherent flexibility, user-friendliness, and heightened security measures foster a conducive environment for streamlined payment logics, potentially revolutionizing payment structures in the DeFi space.